# Non-destructive testing of carbon reinforced plastics by means of phase retrieval

Mostafa Agour (1,2), Claas Falldorf (1), Christoph v. Kopylow (1), Ralf B. Bergmann (1)

(1) Bremer Institut für angewandte Strahltechnik, Klagenfurter Str.2, 28359 Bremen, Germany, agour@bias.de
(2) Physics Department, Aswan Faculty of Science, South Vally University, 81528 Aswan,Egypt

**Presenting author:** Mostafa Agour

**Presentation preference:** oral

Optical non-destructive testing methods (NDT) have been widely used for detecting failures and defects hidden below investigated surfaces. Most established methods are speckle based techniques e.g., Holographic Interferometry [1], Electronic Speckle Pattern Interferometry (ESPI) [2] and Shearography [3]. They are based on the superposition principle and the inspection process requires two sequential measurements. The first measurement is recording the intensity of superposing the wave field scattered from the area under investigation and a reference wave. This is followed by a second measurement after applying a stress e.g. applying vacuum suction, pressurization or creating a thermal stress. Because these effects produce a displacement or strain, one may then conclude on the hidden defects by subtracting the phase distribution extracted from both measurements. However, the main disadvantages of these techniques arising from the use of the superposition principle which raises high demands regarding coherence of light used.

Recently, we have developed a new inspection method by means of phase retrieval [4, 5]. It is based on a phase-only spatial light modulator (SLM) located across the Fourier domain of a 4f-imaging system [6, 7]. Within this approach the SLM modifies electronically the wave field by means of the transfer function of free space propagation, which is a pure phase function. Accordingly, a set of intensity measurements associated to different propagated states can be captured across a common camera plane. The advantages of the proposed approach are summarized as follows:
- The major benefit in regards to the state of the art in NDT is that no reference beam is required during the recording process.
- In contrast to existing phase retrieval approaches, the measuring time is considerably reduced due to the chosen setup.

This enables NDT by means of phase retrieval from a set of intensity observations [4].

In this work, the developed system will be used to inspect carbon reinforced plastics samples (CFRP) under applying a thermal load. For this purpose, the system is used to capture a sequence of 8 spatially separated recording planes, where the distance between subsequent planes equals 2 mm. For detecting the hidden failures two sets of intensity observations are recorded. The first set for the initial state and the second set is captured after applying the load.  To recover the phase information's associated with the two states, the captured intensities have been subjected to an iterative algorithm based on the method of generalized projection [8].